# Artificial intelligence-enabled precision medicine for inflammatory skin diseases

**Authors:** Alice Tang, PhD[1], Maria L. Wei, MD, PhD[2,3], Anna Haemel, MD[2], Cindy La, MD[4], Marina Sirota, PhD[1], Ernest Y. Lee, MD, PhD[*1,2]

**Affiliation**
[1] Bakar Computational Health Sciences Institute, University of California, San Francisco, San Francisco, CA
[2] Department of Dermatology, University of California, San Francisco, San Francisco, California, USA.
[3] Dermatology Service, San Francisco VA Health Care System
[4] Department of Internal Medicine, Southern California Permanente Medical Group, Baldwin Park, California, USA

**Corresponding author**
Ernest Y. Lee, MD, PhD
Bakar Computational Health Sciences Institute, University of California, San Francisco, San Francisco, CA, Department of Dermatology, University of California, San Francisco, San Francisco, California, USA.
ernest.lee@ucsf.edu

**Abstract**

Recent advances in artificial intelligence (AI) and multimodal data collection are revolutionizing dermatology. Generative AI and machine learning approaches offer opportunities to enhance the diagnosis and treatment of inflammatory skin diseases, including atopic dermatitis, psoriasis, hidradenitis suppurativa, and autoimmune connective tissue disease. This review examines the current landscape of AI applications for inflammatory skin diseases and explores how generative AI and machine learning methods can advance the field through deep phenotyping, disease heterogeneity characterization, drug development, personalized medicine, and clinical care. We discuss the promises and challenges of these technologies and present a vision for their integration into clinical practice.

**Introduction**

The increasing availability of comprehensive patient information, including multimodal imaging, multiomic profiles (genomics, transcriptomics, proteomics, epigenomics, metabolomics, etc.), and detailed clinical information, is driving a significant transformation in healthcare. Artificial intelligence (AI), an umbrella term for computational methods that mimic human intelligence, has enabled advances in disease diagnosis, treatment planning, and clinical decision support (Khera et al. 2023). Within AI, machine learning (ML) algorithms can learn from data to identify patterns or make predictions, while generative models can simulate data or mimic human reasoning. ML approaches can be broadly categorized into supervised learning, where models learn from labeled data, and semi-supervised or unsupervised learning, where models identify structures using partially labeled or unlabeled data. Dermatology, with its strong reliance on visual cues and its developing integration of diverse data modalities, offers fertile ground for the application of these AI technologies.

The application of AI, specifically deep convolutional neural networks (CNNs), for the image classification of skin lesions, including melanoma and other skin cancers, has emerged as a classic example in recent decades. A prominent demonstration of this was in 2017, when deep CNNs achieved performance comparable to dermatologists in classifying such lesions (Esteva et al. 2017). Further applications beyond classification of skin cancer and melanoma have since been developed, including model improvements, dermatopathology classification, teledermatology triaging, and beyond (Cazzato and Rongioletti 2024; Muhaba et al. 2022; Soenksen et al. 2021).

Within the broader landscape of AI, generative models and machine learning methods have emerged as particularly promising tools for advancing personalized medicine in dermatology. These technologies extend beyond traditional supervised classification tasks: generative AI can create synthetic data and bridge communication gaps for providers and patients, while unsupervised learning excels at discovering hidden patterns in complex diseases without predetermined classifications. These capabilities are especially valuable for inflammatory skin diseases, where complex pathophysiology and disease heterogeneity pose substantial challenges in diagnosis, classification, and treatment optimization. The ability to accurately characterize this heterogeneity across diverse patient populations and skin types is essential for advancing personalized medicine approaches that precisely target immune mechanisms.

This comprehensive scoping review will encompass current applications of machine learning and AI in dermatology, with a specific focus on inflammatory skin disorders (**Figure 1**). Conditions discussed include but are not limited to atopic dermatitis, psoriasis, hidradenitis suppurativa, morphea/scleroderma, vitiligo, alopecia, acne vulgaris, and rosacea (**Table 1**). We will explore and synthesize the literature spanning applications of AI in disease classification and prediction, deep phenotyping, biomarker discovery, therapeutic design, and clinical care (**Figure 1** and **Supplemental Information**). We will also explore current challenges and opportunities for future diagnostic and therapeutic development and their potential for clinical implementation.

## Current Applications in Inflammatory Skin Diseases

### Classification and Disease Prediction

Given the inherent visual nature of dermatology, artificial intelligence in dermatology has flourished along with advances in computer vision and image analysis, incorporating methods such as convolutional neural networks and vision transformers. While early AI applications and advancements have primarily focused on melanoma detection (Choy et al. 2023; Esteva et al. 2017), recent developments have expanded to encompass a greater diversity of skin diseases, including chronic inflammatory skin disorders such as acne, psoriasis, eczema, rosacea, vitiligo, with multiple publications focused on single-disease or multi-class classification (Choy et al. 2023; Li Pomi et al. 2024) (**Table 1**, first column). One group trained AI models to help distinguish between conditions that may look similar for inexperienced physicians, such as rosacea, from other facial inflammatory diseases (Ge et al. 2022).

Beyond traditional clinical images, the scope of classification and prediction has also been expanded to accommodate a greater heterogeneity of input data sources, from histological slides (Hosny et al. 2020; Muhaba et al. 2022) to non-image data such as gene expression profile sequencing from tape strips, Raman spectroscopy, and electronic health records for atopic dermatitis and psoriasis (Greenfield et al. 2023; Gustafson et al. 2017; He et al. 2021; Jiang et al. 2022) (**Figure 1**, top row). This versatility enables comprehensive analysis across multiple domains and processes in the approach to inflammatory skin disorders, including low-level processing tasks such as lesion segmentation, to higher-level tasks such as differentiating similar diagnoses and recommending treatments (Hammad et al. 2023; Li et al. 2023; Yu et al. 2020).

A particularly promising development is the emergence of interpretable AI models, including approaches that probe into attention networks and gradient-based saliency maps (Anjali et al. 2025; Zhong et al. 2024). These methods enhance our understanding of specific features, such as morphology or pigmentation, that influence model decisions for conditions like rosacea or vitiligo (Anjali et al. 2025; Mohan et al. 2025; Zhong et al. 2024). Interpretability helps provide transparency and allows for scrutiny of AI-driven outputs, which is essential for fostering trust in clinical applications. Further advances include the incorporation of transfer learning and the fine-tuning of existing models, or the specialized adaptation of pre-trained models on a smaller or task-specific dataset. This allows for efficient training and effective learning by leveraging learned model parameters and knowledge gained from broad tasks (e.g. ResNet-50) for use on smaller datasets such as shown in chronic plaque psoriasis (Chakraborty et al. 2024; Hosny et al. 2020), reducing the need for extensive new datasets or computational resources.

A few studies exploring transformer architectures represent another promising advancement, enabling enhanced recognition or explainability of disease patterns as well as variations across skin type (Abdel Mawgoud and Posch 2025; Khan and Khan 2023; Mohan et al. 2025; Zhong et al. 2024) (**Table 1**, bottom row). Multimodal data modalities are also increasingly incorporated, from genetic and transcriptomics data, to clinical and wearable data as well (Andreoletti et al. 2021; Hurault et al. 2020; Muhaba et al. 2022). However, there is still a lack of substantial AI research applied to inflammatory skin diseases. While conditions like atopic dermatitis and psoriasis have received considerable attention, other inflammatory conditions remain understudied. For example, there is an unmet need for approaches to distinguish inflammatory conditions from infectious and malignant processes, particularly in cases where clinical presentations may overlap (Ujiie et al. 2022).

**Phenotyping, Subtyping, Biomarkers, and Treatments**
In addition to improving our diagnostic approach, machine learning methods in dermatology can serve to expand our understanding of inflammatory skin diseases, enabling applications such as disease characterization, biological hypothesis generation, and precision medicine approaches (Félix Garza et al. 2019; Tang et al. 2024). Biomarker identification and mechanistic insights can be further elucidated with evolving computational approaches that integrate diverse data types - from genomics, transcriptomics, cell surface markers, and clinical phenotypes, enabling understanding of patterns and relationships across modalities. For example, one study identified interferon-stimulated gene 15 as a dermatomyositis biomarker

(Wang et al. 2024c), and another identified the role of keratinocyte immunophenotype in drug response in atopic dermatitis (Clayton et al. 2021).

Unsupervised learning methods such as embedding and clustering algorithms enable the characterization of disease heterogeneity by identifying subtypes based on complex patterns in the data instead of preconceived classifications. For example, machine learning on transcriptomics data has revealed distinct subtypes in atopic dermatitis, psoriasis, and juvenile dermatomyositis, that may reflect unique molecular mechanisms, treatment responses, and prognoses (Berna et al. 2020; Neely et al. 2022; Zhang et al. 2024a) (**Table 1**, second column). Some of these studies utilized interpretable models or interpretation analysis (e.g. grad-CAM) for supervised or unsupervised learning. This can enable discovery sciences for identification of key disease characteristics such as visible morphological characteristics, molecular expression differences, or potential therapeutic development (**Figure 1**, examples in Phenotyping, Subtyping, & Personalized Treatment row).

Multiple data modalities that combine molecular profiles with clinical data can also enhance our understanding of disease across diverse populations. This methodology has been especially valuable in complex conditions like systemic lupus erythematosus, scleroderma, morphea, and other autoimmune disorders, where target inflammatory pathways and subtypes have been identified that can help inform treatment (Andreoletti et al. 2021; Choi et al. 2023; Cutts et al. 2024; Franks et al. 2019). Subtypes may reflect underlying patient or molecular heterogeneity, or disease severity for prognosis. For example, studies have identified risk of therapeutic resistance in atopic dermatitis or risk of alopecia universalis from alopecia areata (Wu et al. 2022a; Zhang and Nie 2022).

With regards to treatment advances, efforts are underway to study the integration of protein-drug interaction networks that can further accelerate therapeutic target identification and drug repurposing efforts (Patrick et al. 2018a). Mathematical models of biological interactions are another approach to increase the understanding of biological processes and drug response (Barraza et al. 2024; Chen et al. 2021; Kardynska et al. 2023). Scientists from Absci have demonstrated the potential for generative AI to design antibodies against targets of interest for the treatment for alopecia and other inflammatory diseases (Absci). These approaches analyze drug response pathways to predict treatment efficacy for specific disease subtypes, while transcriptomic analyses can provide deeper insight into disease mechanisms as another approach to identify targets for drug development (Yamanaka et al. 2023) (**Table 1**, second column).

Moreover, integration with population-level information can elucidate risks from external exposures, given the skin's interaction with the environment (e.g. pollution, humidity), as well additive risks due to comorbidities (e.g. diabetes mellitus, mental health disorders) as shown in studies that look at external or comorbid factors influencing irritant dermatitis and hidradenitis suppurativa (Fortino et al. 2020; Hua et al. 2021; Huang et al. 2021b; Papa et al. 2023). AI approaches are also investigated at the stage of clinical trial design and treatment selection through patient stratification based on disease severity, subtypes, or predicted responses (Huang et al. 2023; Hurault et al. 2020). The integration of supervised and unsupervised learning methods, alongside diverse data types, enables the discovery of complex relationships among biological mechanisms, environmental exposures, and clinical manifestations. This comprehensive analytical approach enhances the understanding of disease processes and supports the advancement of precision medicine strategies. These applications are still at the beginning stages, with opportunities for further expansion.

**Advancing Clinical Workflows and Patient Care with Generative AI**
The application of generative AI is beginning to reshape clinical workflows across medicine, and new uses in dermatology are emerging. Many recent advances are focused on applications across medical practice and across specialties. For instance, the development of AI scribes and clinical summarizers enhances clinical efficiency by automating documentation and supporting workflows (Miao et al. 2025; Williams et al. 2024). Foundation models are also driving advanced multimodal analysis by integrating visual data with clinical information and medical literature to support clinical decision-making with detailed differential diagnoses and tailored recommendations (Kim et al. 2024; McDuff et al. 2025). Furthermore, generative AI can improve patient communication by aiding in translation or creating accessible explanations of medical information (Aydin et al. 2024; Johri et al. 2025). Beyond patient care, the potential for generative AI has also been proposed for applications such as biomarker identification and drug design (Shanehsazzadeh et al. 2023; Ying et al. 2024). The generative models enabling these advancements vary widely, encompassing approaches from foundational physical equations and probability distributions to sophisticated neural network architectures such as transformers (including large language models), generative adversarial networks, and diffusion models (**Figure 1**, left column).

Within dermatology specifically, initial generative AI efforts focused on image enhancement and data augmentation, often to aid image-based classification models to address dataset limitations such as low sample size and artifacts such as hair and ink markings. The goals have evolved further to overcome limitations in dataset diversity, image acquisition bias, and other confounding factors that can influence the quality of datasets (**Figure 1**, bottom row, augmentation and simulation). For example, the DermDiff model aims to mitigate biases in image datasets by simulating various skin tones (Munia and Imran 2025), although bias amplification can still arise with image generation models (Mikołajczyk et al. 2022). Vision transformer models are also being developed to generate explanations or interpretations of dermatological images directly (Lin et al. 2025; Mohan et al. 2025; Zhou et al. 2023).

Prototype studies, primarily in skin cancer, have demonstrated the potential of generative AI to aid diagnostic decision-making for both dermatologists and non-specialists, providing valuable guidance for clinical assessment and training. For instance, one study explored integrating generative AI with ABCDE rule analysis for enhanced melanoma diagnosis and dermatology education (Jütte et al. 2024). Similarly, AI-generated translations of dermatopathology reports aim to make complex information more understandable for patients (Zhang et al. 2024b). The accelerated adoption of telehealth after the COVID-19 pandemic also enhanced teledermatology advancements, spurring investigations into how generative AI can aid with virtual consultations, enhance remote image assessment, and facilitate accurate documentation (Shapiro and Lyakhovitsky 2024; Vodrahalli et al. 2023; Vodrahalli et al. 2021) (**Figure 1**, bottom row). This technological advancement holds significant promise for expanding access to dermatological expertise, especially for patients in underserved or remote areas (Muñoz-López et al. 2021).

While these generative AI applications in dermatology are promising, applications in inflammatory skin diseases remain an area with significant gaps (**Table 1**, right column). A couple of studies are emerging that investigate the potential for generative AI to enhance diagnostic support, for example, in lupus erythematosus diagnosis or alopecia areata severity evaluation (Lee et al. 2020; Li et al. 2024). Studies have also evaluated the utility of large language models in answering questions from patients regarding psoriasis and acne vulgaris (Kılıçoğlu et al. 2025; Lakdawala et al. 2023). Generative AI has also demonstrated promise in antibody design for diseases like alopecia areata, but these developments are still in early stages (Absci). While foundational capabilities of generative AI are being established in medicine and general dermatology, dedicated research and development will be needed to harness its potential for complex challenges provided by inflammatory skin diseases.

**Challenges, Opportunities, and Future Directions**

In summary, artificial intelligence is rapidly transforming dermatology by providing novel tools to address the complex diagnostic and therapeutic challenges associated with skin diseases. While early AI applications were predominantly focused on skin cancer detection, more recent advances in AI have expanded applications to include inflammatory skin diseases, including atopic dermatitis, psoriasis, hidradenitis suppurativa, and autoimmune connective tissue disorders. The emergence of multimodal data integration, ranging from clinical images and histopathology to transcriptomic and genomic profiles, has enabled more refined classification, disease sub-phenotyping, and identification of novel biomarkers and therapeutic targets (**Figure 1**). These insights can support the development of precision medicine approaches that tailor treatments to individual molecular profiles and clinical characteristics.

Looking ahead, the convergence of multimodal data analysis offers unprecedented opportunities for precision medicine in dermatology. By moving beyond images and integrating data such as genetic profiles, transcriptomics, exposomics, clinical information, and patient outcomes, AI systems can be utilized to match patients with optimal therapies based on individual disease pathways and subphenotypes (Patrick et al. 2018a; Wongvibulsin et al. 2022). This approach could significantly reduce the trial-and-error nature of current treatment selection for possibly poorly defined diagnoses of inflammatory skin conditions. Additionally, AI-powered analysis of population-level data could identify communities at elevated risk for developing inflammatory conditions, enabling earlier interventions and preventive strategies.

The integration of generative AI and machine learning in dermatology presents both significant opportunities and important challenges for basic scientists and clinicians. Several key challenges persist. AI applications in inflammatory skin diseases beyond psoriasis and atopic dermatitis remain relatively understudied, such as in vitiligo and hidradenitis suppurativa. Algorithmic bias, particularly related to the underrepresentation of skin of color, continues to limit the generalizability of AI models. AI approaches inherently also risk exacerbating health inequities and magnifying institutional racism (Grzybowski et al. 2024). Misinformation is also rampant, with significant bias, inconsistent validation, and inadequate privacy protection (Wongvibulsin et al. 2024). Key technical hurdles include standardizing image quality across photo acquisition methods, addressing confounding factors and artifacts in images, and developing representative datasets that span diverse patient populations (Choy et al. 2023; Daneshjou et al. 2022), although preliminary models are evolving that can provide feedback on image quality during acquisition (Vodrahalli et al. 2023). Additionally, the clinical translation of many AI tools remains in its infancy due to limitations in data quality, external validation, and integration into existing workflows. Successful clinical implementation requires careful attention to workflow integration, provider education, and patient acceptance, to ensure that AI tools can enhance care delivery worldwide.

The future role of AI in dermatology will eventually include the integration of generative AI tools that support both routine clinical care and complex decision-making. Generative AI models will enhance medical education, clinical simulation, and patient communication (Motolese et al. 2022; Zhang et al. 2024b), while improved analysis of molecular and cellular mechanisms will enable the development of targeted treatments. Advancing the management of inflammatory skin disorders will require sustained collaboration among clinicians, researchers, pharmaceutical experts, and technology developers (Sengupta 2023), with a strong focus on ethical considerations and equitable access to AI-enhanced care

**Figure 1: Landscape of current and future applications of artificial intelligence to advance the precision medicine of inflammatory skin diseases.**

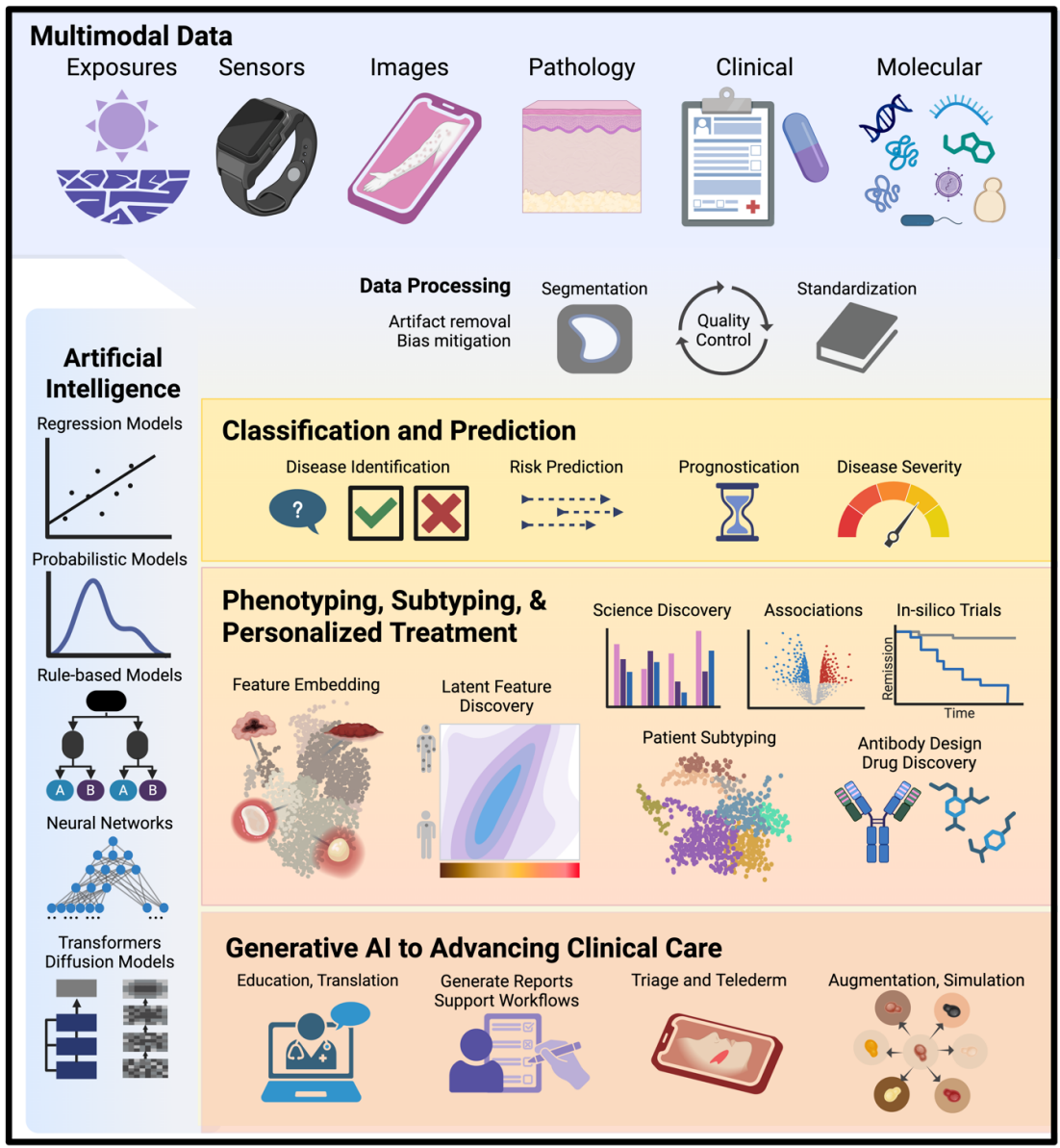

The figure summarizes the landscape of AI in inflammatory dermatology. First there is data from external and internal factors encompassing various modalities shown in the top row. The raw data may undergo preprocessing, such as segmentation and quality control, to prepare for subsequent tasks. The left column reveals the range of AI approaches and algorithms for various applications in the figure. Classification and prediction utilize known labels or outcomes for a specific task, such as diagnosis or prognosis. Exploratory tasks may include identifying features (or an attribute) that is directly quantifiable (e.g. area of lesion) or indirectly quantified (e.g. latent features). This may then lead to improvement in a supervised ML task, or aid in discovery tasks (e.g. cell surface marker associations) and precision medicine approaches (e.g.

patient subtyping). In silico means through computer modelling or simulation. The bottom rows show how generative AI can integrate with research and clinical care.

**Table 1: AI-enabled advancements across the spectrum of inflammatory skin diseases**

Reviewed studies grouped by skin disorder in rows or grouped by goals in columns

| Diseases | Approach or Application | | |
|---|---|---|---|
| | **Classification & Disease Prediction** | **Phenotyping, Subtyping, Biomarkers, & Treatments** | **Advancing Clinical Workflows & Patient Care with Generative AI** |
| **Atopic Dermatitis** | Dev et al. 2022; Fortino et al. 2020; Ghosh et al. 2015; Hammad et al. 2023; Ho et al. 2020; Huang et al. 2021b; Hurault et al. 2020; Jiang et al. 2022; Liu et al. 2020; Muhaba et al. 2022; Muñoz-López et al. 2021; Wu et al. 2020b; Zhu et al. 2021b | Berna et al. 2020; Clayton et al. 2021; Dev et al. 2022; Fortino et al. 2020; Greenfield et al. 2023; Gustafson et al. 2017; Jiang et al. 2022; Khan et al. 2024; Maintz et al. 2021; Martínez et al. 2022; Prabhu et al. 2018; Prakash et al. 2020; Wang et al. 2024b; Wang et al. 2022; Wu et al. 2023; Wu et al. 2022a | Czajkowska et al. 2022; Czajkowska et al. 2021a; Fortino et al. 2020; Giavina Bianchi et al. 2025; Ho et al. 2020; Huang et al. 2021b; Khan et al. 2024; Lakdawala et al. 2023; Maintz et al. 2021; Prakash et al. 2020; Prasannanjaneyulu et al. 2022; Sulejmani et al. 2024; Yang et al. 2025; Zhang et al. 2024b |
| **Psoriasis** | Aijaz et al. 2022; Chakraborty et al. 2024; Eskandari and Sharbatdar 2024; Hammad et al. 2023; Huang et al. 2021a; Hurault et al. 2020; Karthik et al. 2022; Liu et al. 2020; Shrivastava et al. 2016; Wu et al. 2020b; Yu et al. 2022; Zhao et al. 2020; Zhou et al. 2024; Zhu et al. 2021b | Aijaz et al. 2022; Damiani et al. 2020; Martínez et al. 2022; Moon et al. 2024; Okamoto et al. 2022; Patrick et al. 2018b; Patrick et al. 2018a; Pournara et al. 2021; Shrivastava et al. 2017; Shrivastava et al. 2016; Tomalin et al. 2020; Xu and Zhang 2017; Zhang et al. 2024a; Zhou et al. 2024 | Czajkowska et al. 2022; Czajkowska et al. 2021b; Kılıçoğlu et al. 2025; Lin et al. 2025; Moon et al. 2024; Okamoto et al. 2022; Queiro et al. 2022; Schaap et al. 2022; Tomalin et al. 2020; Xu and Zhang 2017; Yu et al. 2020; Zhang et al. 2024a; Zhang et al. 2024b |
| **Vitiligo** | Guo et al. 2022; Zhong et al. 2024 | Hillmer et al. 2024; Wang et al. 2021a | Hillmer et al. 2024 |
| **Alopecia** | Liu et al. 2020; Muñoz-López et al. 2021; Zhang and Nie 2022 | Absci; Chen et al. 2021; Xiong et al. 2023; Zhang and Nie 2022 | Absci; Lee et al. 2020; Shapiro and Lyakhovitsky 2024; Zhang et al. 2024b |
| **Acne** | Karthik et al. 2022; Liu et al. 2020; Muhaba et al. 2022; Muñoz-López et al. 2021 | Abdel Mawgoud and Posch 2025; Liu et al. 2022a; Yang et al. 2021 | Abdel Mawgoud and Posch 2025; Lakdawala et al. 2023; Yang et al. 2021 |
| **Hidradenitis Suppurativa** | Ali et al. 2025; Kirby et al. 2024; Muñoz-López et al. 2021 | Cazzaniga et al. 2021; González-Manso et al. 2021; Wiala et al. 2024 | |
| **Rosacea** | Ge et al. 2022; Park et al. 2023; Zhao et al. 2021; Zhu et al. 2021b | Anjali et al. 2025; Barraza et al. 2024; Binol et al. 2020; Deng et al. 2023; Gao et al. 2025; Ge et al. 2022; Mao and Li 2024; Nicholas et al. 2024; Rajalingam et al. 2023; Wang et al. 2024a | Barraza et al. 2024; Nicholas et al. 2024; Rajalingam et al. 2023 |
| **Cutaneous Lupus Erythematosus** | Li et al. 2024; Nair et al. 2025; Wu et al. 2021 | Dunlap et al. 2022; Martínez et al. 2022; Perez et al. 2022; Tao et al. 2025; Zhu et al. 2021a | |
| **Morphea or Scleroderma** | Huang et al. 2021a | Abdel Mawgoud and Posch 2025; Choi et al. 2023; Franks et al. 2019; Martínez et al. 2022 | Abdel Mawgoud and Posch 2025 |
| **Infectious Skin Disease** | Barbieri et al. 2022; Muñoz-López et al. 2021; Shen et al. 2024; Thieme et al. 2023 | Barbieri et al. 2022; Koo et al. 2021; Thieme et al. 2023 | Hutchinson et al. 2023; Thieme et al. 2023 |
| **Other or Multiple** | 3Derm Systems, Inc; Ahmad et al. 2022; Capurro et al. 2024; Corbin and Marques 2023; Czajkowska et al. 2021b; Esteva et al. 2017; Huang et al. 2021a; Lin et al. 2025; Liu et al. 2022b; Liu et al. 2020; Mohan et al. 2025; Muhaba et al. 2022; Muñoz-López et al. 2021; Shen et al. 2024; Wang et al. 2021b; Wu et al. 2022b; Zhang et al. 2024c; Zhou et al. 2023; Zhu et al. 2021b | Abdel Mawgoud and Posch 2025; Abhishek et al. 2021; Bettuzzi et al. 2022; Esteva et al. 2017; Hart et al. 2024; Koopman et al. 2024; Liu et al. 2020; McLeish et al. 2023; Mikołajczyk et al. 2022; Neely et al. 2022; Patrick et al. 2018a; Prabhu et al. 2018; Shanehsazzadeh et al. 2023; Shen et al. 2024; Wang et al. 2024c; Wang et al. 2021b; Wu et al. 2020a; Xue et al. 2023; Ying et al. 2024; Zhu et al. 2021b | 3Derm Systems, Inc; Abdel Mawgoud and Posch 2025; Abhishek et al. 2021; Czajkowska et al. 2022; Czajkowska et al. 2021a; Groh et al. 2024; Jütte et al. 2024; Kim et al. 2024; Koopman et al. 2024; Lin et al. 2025; Mikołajczyk et al. 2022; Muhaba et al. 2022; Munia and Imran 2025; Muñoz-López et al. 2021; Shanehsazzadeh et al. 2023; Vodrahalli et al. 2021; Zhang et al. 2024b; Zhang et al. 2024c; Zhou et al. 2023 |

**ORCIDs:**
Alice Tang - 0000-0003-4745-0714
Maria Wei - 0000-0002-3568-1921
Marina Sirota - 0000-0002-7246-6083
Ernest Y. Lee - 0000-0001-5144-2552

**Conflict of Interest:**
All other authors declare no competing interests.

**Acknowledgements:**
A.S.T is supported by the UCSF Medical Scientist Training Program T32GM007618 and F30 Fellowship 1F30AG079504. M.W. is supported by DoD grant W81XWH2110982.
M.S. is supported by NIHT grant P30AR070155. E.Y.L is supported by NIH grant T32AR007175, a Dermatology Foundation Dermatologist Investigator Research Fellowship and Career Development Award, and a Hidradenitis Suppurativa Foundation Translational Research Grant.

**Author Contributions:**
Conceptualization: A.T., A.H., E.L.; Figure and Table: A.T; Writing – original draft: A.T.; Writing - review: A.T., A.H., M.L., C.L., M.S., E.L.

**Supplementary Materials:**
Supplement: Individual Paper Descriptions